\documentclass[12pt]{article}

\usepackage{amsmath,amssymb,amsbsy,amsfonts,amsthm,latexsym,
                        amsopn,amstext,amsxtra,euscript,amscd,color}
\usepackage{color,xcolor}
\usepackage{latexsym}
\usepackage{cite}
\usepackage{amsfonts}
\usepackage{amsfonts,mathrsfs}
\usepackage{graphicx}
\usepackage{psfrag}
\usepackage{subfigure}
\usepackage{url}
\usepackage{stfloats}
\usepackage{amsmath}

\newtheorem{theorem}{Theorem}
\newtheorem{lemma}{Lemma}
\newtheorem{corollary}{Corollary}

\newcommand{\quash}[1]{}

\begin{document}

\title{Linear code derived from the primage of quadratic function}

\author{Xiaoni Du$^{1}$, Yunqi Wan$^{1}$\\
1. College of Mathematics and Information Science, \\Northwest Normal University, \\ Lanzhou, Gansu
730070, P.R. China\\
ymLdxn@126.com\\
}

\maketitle

\begin{abstract}
Linear codes have been an interesting topic in both theory and practice for many years. In this paper, for an odd prime power $q$, we construct some class of linear code over finite field $\mathbb{F}_q$ with defining set be the preimage of general quadratic form function and determine the explicit complete weight enumerators of the linear codes. Our construction cover all the corresponding result with quadratic form function and they may have applications in cryptography and secret sharing schemes.

\end{abstract}

\section{Introduction}

Let $p$ be an odd prime, $m,~e$ be positive integers and $q=p^e$. An $[n,k,d]$ linear
code $\mathcal{C}$ over the finite field $\mathbb{F}_q$ is a $k$-dimensional subspace of $\mathbb{F}_q^n$ with minimum (Hamming) distance $d$, which determines the error correcting capability of $\mathcal{C}$. Let $A_i$ denote the number of codewords with Hamming weight $i$ in a code $\mathcal{C}$ of length $n$. The weight enumerator of $\mathcal{C}$ is defined by
$$1+A_1z+A_2z^2+\ldots+A_nz^n.$$
The sequence $(A_1,~ A_2, \ldots, A_n)$ is called the weight distribution of $\mathcal{C}$. The code $\mathcal{C}$ is called to be $t$-weight if the number of nonzero $A_j~(1 \le j \le n)$ in the sequence
$(A_1,~ A_2, \ldots, A_n)$ equals to $t.$

The complete weight enumerator of a code $\mathcal{C}$ over $\mathbb{F}_q$ enumerates the codewords according to the number of symbols of each kind contained in each codeword. Denote $\mathbb{F}_q=\{\omega_0, \omega_1, \cdots, \omega_{q-1}\}$ by the finite elements with $\omega_0=0$ and $\mathbb{F}_q\backslash\{0\}$ by $\mathbb{F}^*_q$, respectively. For a codeword $\mathbf{c}=(c_0, c_1, \cdots, c_{n-1})\in \mathbb{F}^n_q$, The weight weight enumerator $\omega [\mathbf{c}]$ of $\mathbf{c}$ is defined by $$\omega[\mathbf{c}]=\omega^{k_0}_0\omega^{k_1}_1\cdots\omega^{k_{p-1}}_{p-1}, $$ where $\sum\limits_{j=0}^{p-1}k_j=n$, $k_j$ is the number of components of $\mathbf{c}$ equal to $\omega_j$. The complete weight enumerator of the code $\mathcal{C}$ is then \[CWE(\mathcal{C})=\sum\limits_{\mathbf{c}\in \mathcal{C}}\omega[\mathbf{c}].\]

The weight distribution gives the minimum distance of the code, and hence the error correcting capability. Furthermore, the weight distribution of a code allows the computation of the error
probability of error detection and correction with respect to some error detection and error correction algorithms (see \cite{Klove-95} for details). Thus the study of the weight distribution attracts
much attention in coding theory and much work focus on the determination of the weight distributions of linear codes (see \cite{Ding-15, Ding-Liu-Ma, Ding-yang-13, Ding-Ding, Feng-Luo, Xu-Cao, Zhou-Ding-13, Zhou-Ding-14, Zhou-li-fan-Helleseth}
and the references therein). Linear codes can be applied in consumer electronics, communication and data storage system. Linear codes with few weights are of important in
secret sharing \cite{Carlet-Ding-Yuan-05,Yuan-Ding-05}, authentication codes \cite{Ding-wang-TCS-05}, association schemes \cite{Calderbank-Goethals-84} and strongly regular graphs \cite{Calderbank-Kantor-86}.

It is easy to see that the complete weight enumerators for binary linear code are just the weight enumerators, while for nonbinary case, the weight enumerators can be obtained from their complete weight enumerators.
Furthermore, the complete weight enumerators are closely related to the deception of some authentication codes constructed from linear codes \cite{Dt07}, and used to compute the Walsh transform of monomial functions over finite fields \cite{Helleseth-Kholosha-06}. Thus, a great deal of research \cite{bk91,ccd06,d08,dy06,k89,kn99,ly15,yy159,yy15} is devoted to the computation of the complete weight distribution of
specific codes over $\mathbb{F}_p$.

Let $D = \{d_1, d_2, \ldots, d_n\} \subseteq \mathbb{F}_{q^m}$. Then a linear code of
length $n$ defined over $\mathbb{F}_q$ is $$\mathcal{C}_D = \{(Tr_1^m(xd_1),Tr_1^m(xd_2), \ldots, Tr_1^m(xd_n)):~ x\in \mathbb{F}_{q^m} \}$$
where $Tr^m_1$ is the trace function from $\mathbb{F}_{q^m}$ to $\mathbb{F}_q$. The set $D$ is called the defining set of $\mathcal{C}_D$. This construction method
is used for obtaining linear codes with few weights \cite{Ding-Ding, Zhou-li-fan-Helleseth}. The selection of $D$ directly affects the constructed linear
codes. How to choose defining sets for good linear codes is interesting and important and thus many linear codes can be obtained  from the proper selection of $D$.

Let $F(x)\in\mathbb{F}_q[x]$ and
$f(x) = Tr_1^m(F(x)).$ Most of the the previous works
focus on changing or generalizing the function $F(x)$ or $f(x),$
i.e, square functions, quadratic bent functions
$f(x)$ \cite{Zhou-li-fan-Helleseth}, and weakly regular bent functions \cite{Tang-Li-Qi-Zhou-Helleseth} with $D$ be the kernel of $f(x)$, i.e., $D=\{x\in \mathbb{F}_{q^m}^*:~f(x)=0 \}$ with $e=1$.

Very recently, Zhou \emph{et al.} construct some classes of linear code over $\mathbb{F}_{p}$ with defining set be the kernel of the
quadratic form function of full rank (quadratic Bent function) and examined their weight distribution \cite{Zhou-li-fan-Helleseth}. Later, Zhang \emph{et al.} extend Zhou \emph{et al.}'s work to $\mathbb{F}_{q}$ with defining set be the kernel of general quadratic form function and examined the complete weight enumerators of the codes\cite{Zhang-Zhou-peng}. In this manuscript, we construct some class of linear code over over $\mathbb{F}_{q}$ with defining set be the preimage of general quadratic form function and examined the complete weight enumerators of the codes and discussed their application in secret sharing schemes.

\section{Quadratic form function}

Identifying $\mathbb{F}_{q^m}$ with the $m$-dimensional $\mathbb{F}_q$-vector space $\mathbb{F}_q^m$, a function $Q(x)$ from $\mathbb{F}_{q^m}$ to $\mathbb{F}_q$ can be regarded as an $m$-variable polynomial over $\mathbb{F}_q$. The former is called a quadratic form over $\mathbb{F}_q$ if the latter is a homogeneous polynomial of degree two in the form
$$Q(x_1,x_2,\ldots, x_m)= \sum\limits_{1\le i\le j \le m} a_{ij}x_ix_j, $$
where $a_{ij}\in\mathbb{F}_q $. Any choice of a basis $\{\beta_1, \beta_2, \ldots, \beta_m\}$ from $\mathbb{F}_{q^m}$ as a vector space over $\mathbb{F}_q$ determines an identification $\mathbb{F}_{q}^m \rightarrow \mathbb{F}_{q^m}$ by $\bar{x}=(x_1,x_2,\ldots, x_m) \mapsto \sum_{i=1}^m x_i\beta_i=x$. We write $\bar{x}$ when
an element is to be viewed as a vector in $\mathbb{F}_{q}^m$, and we write $x$ when the same vector is to be viewed as an element of $\mathbb{F}_{q^m}$.
The rank of the quadratic form $Q(x)$ is defined as the codimension of the $\mathbb{F}_q$ -vector space
$$V = \{y\in \mathbb{F}_{q^m}:~ Q(x+y)-Q(x)-Q(y) = 0 \mathrm{~for~ all~} x\in \mathbb{F}_{q^m} \}.$$
That is $|V| = q^{m-r}$ where $r$ is the rank of $Q(x).$

In the sequel, we shall give some lemmas that will be needed for our main results. Before doing this, we first fix some notation.

\begin{enumerate}
\item[*] $B_{2j}(\bar{x}) = x_1x_2 +x_3x_4+\ldots+x_{2j-1}x_{2j}$ where $j$ is an integer with $0\le j \le m$ (we assume
that $B_0 = 0$ when $j = 0$).\\
\item[*] $I(x)$ is a function over $\mathbb{F}_q$ defined by $I(x) =-1$ for any $x \in \mathbb{F}_{q}^*$ and $I(0) = q-1$.\\
\item[*] $\eta(x)$ is the quadratic character of $\mathbb{F}_q$ with $\eta(0)=0$.
\end{enumerate}

Quadratic forms have been well studied (see \cite{Klapper-93}, \cite{Klapper-95}, \cite{Lidl}, for example). Here we follow the treatment in \cite{Klapper-93} and \cite{Klapper-95}. It should be noted that the rank of a quadratic form over $\mathbb{F}_q$ is the smallest number of variables required to represent the quadratic form, up to nonsingular coordinate transformations. Mathematically, any quadratic form of rank $r$ can be transferred to three canonical forms as follows.

\begin{lemma}\label{klapper-qf-1} (\cite{Klapper-95}) Let $Q(x)$ be a quadratic form over $\mathbb{F}_q$ of rank $r$ in $m$ variables. Under a
nonsingular change of coordinates, $Q(x)$ is equivalent to one of the following three standard types in Table \ref{table1}:
\begin{table}
\begin{center}
\caption{Standard types of quadratic form over $\mathbb{F}_q$ with rank $r$ and $m$ variables}
\label{table1}
\begin{tabular}{ccc}
\hline\noalign{\smallskip} Type of quadratic form  & Parity of $r$ & $N_{\alpha}$ \\
\noalign{\smallskip} \hline \noalign{\smallskip}
I: $B_r(\bar{x})$ &   even &   $ q^{m-1}+I(a)q^{m-r/2-1} $ \\
II: $B_{r-1}(\bar{x})+\mu x_r^2$ & odd  &   $q^{m-1}+\eta(\mu a)q^{m-(r+1)/2}$\\
III: $B_{r-2}(\bar{x})+x_{r-1}^2- \gamma \mu x_r^2$  & even  &   $q^{m-1}-I(a)q^{m-r/2-1}$  \\
\hline
\end{tabular}
\end{center}
\end{table}

where  $\mu \in \{1,\gamma\}$ and $\gamma$ is a fixed nonsquare in $\mathbb{F}_q$ and we denote $N_a$ by
$$N_{a}=\sharp \{\bar{x}\in \mathbb{F}_{q}^m:~Q(\bar{x})=a, ~a \in \mathbb{F}_{q} \}.$$
\end{lemma}

Consider a system of equations consisting of a quadratic form and a linear function. The number of solutions depends on the type and rank of the quadratic form. Let $Q(\bar{x})$ be a
quadratic form of rank $r$ in $m$ variables in one of the three standard types. Let $L_{\bar{b}}(\bar{x})=\bar{b}\bar{x}^T=\sum_{i=1}^m b_ix_i$ be a linear function in $m$ variables with $\bar{b} = (b_1,b_2, \ldots ,b_m) \in \mathbb{F}^m_q \backslash \{\bar{0}\}$, where $ \bar{0}= (0,0, \ldots,0)$ denotes
the all zero vector. For any $a,~v\in \mathbb{F}_q$, we denote by $N(a,v)$ the number of solutions to the system
of equations
$$
\left\{
\begin{array}{ll}
Q(\bar{x})=a\\
L_{\bar{b}}(\bar{x})=v.
\end{array}
\right.
$$

The following lemmas will be used to determine the complete weight enumerators of
linear codes constructed from general quadratic forms over $\mathbb{F}_q$. Before introducing them, we give some notation for the standard quadratic form $Q(\bar{x})$ denoted above. For any vector
$\bar{x} = (x_1, x_2, \ldots, x_m),$ denote $\bar{x'} = (x_1, x_2,\ldots, x_r)$ and $\bar{x}^{''} = (x_{r+1}, x_{r+2}, \ldots, x_m),$ where $r$ is the
rank of $Q(\bar{x}).$ Thus $Q(\bar{x}) = Q(\bar{x'}).$ Let
$$
\hat{Q}(\bar{x})=\left\{
\begin{array}{ll}
Q(\bar{x}), & \mathrm{~if~} Q(\bar{x})=B_r(\bar{x}),\\
B_{r-1}(\bar{x})+\frac{x_r^2}{4 \mu }, & \mathrm{~if~} Q(\bar{x})=B_{r-1}(\bar{x})+\mu x_r^2,\\
B_{r-2}(\bar{x})+\frac{x_{r-1}^2}{4}-\frac{x_{r}^2}{4 \gamma}, & \mathrm{~if~} Q(\bar{x})=B_{r-1}(\bar{x})+x_{r-1}^2-\gamma x_r^2,\\
\end{array}
\right.
$$
where  $\mu \in \{1,\gamma\}$ and $\gamma$ is a fixed nonsquare in $\mathbb{F}_q$. Note that $\hat{Q}(\bar{x})$ is equivalent to $Q(\bar{x})$ under
a change of coordinates.

We always suppose that $\epsilon= 1$ if $Q(x)$ is equivalent to Type I and $\epsilon= -1$ if $Q(x)$ is equivalent to Type III.

\begin{lemma}\label{klapper-2} (\cite{Klapper-95}) Let $Q(\bar{x})$ be a quadratic form over $\mathbb{F}_q$ of rank $r$ in $m$ variables. With notation
as above and $\bar{b}^{''}=0$,

(1) Suppose that $Q$ has Type I or Type III.
$$
N(a,v)=\left\{
\begin{array}{ll}
q^{m-2}+\epsilon I(a)q^{m-1-r/2}, & \mathrm{~if~} Q(\bar{b})=0 \mathrm{~and~} v=0,
\\
q^{m-2}, & \mathrm{~if~} Q(\bar{b})=0 \mathrm{~and~} v\not=0,
\\
q^{m-2}+\epsilon \eta \Big( 4a\hat{Q}(\bar{b})-v^2 \Big)q^{m-1-r/2}, & \mathrm{~if~} Q(\bar{b})\not=0.\\
\end{array}
\right.
$$

(2) Suppose that $Q$ has Type II,
$$
N(a,v)=\left\{
\begin{array}{ll}
q^{m-2}+\eta( \mu a)q^{m-(r+1)/2}, & \mathrm{~if~} Q(\bar{b})=0 \mathrm{~and~} v=0,
\\
q^{m-2}, & \mathrm{~if~} Q(\bar{b})=0 \mathrm{~and~} v\not=0,
\\
q^{m-2}+ I\Big( 4a\hat{Q}(\bar{b})-v^2 \Big) \eta \Big( \mu \hat{Q}(\bar{b}) \Big)q^{m-(r+3)/2}, & \mathrm{~if~} Q(\bar{b})\not=0.\\
\end{array}
\right.
$$
\end{lemma}

\begin{lemma}\label{klapper-3} (\cite{Klapper-95})  Let $Q(\bar{x})$ be a quadratic form over $\mathbb{F}_q$ of rank $r$ in $m$ variables. With notation
as above and $\bar{b}^{''} \not=0$,

(1) Suppose that $Q$ has Type I or Type III.
$$
N(a,v)=q^{m-2}+\epsilon I(a) q^{m-2-r/2},
$$

(2) Suppose that $Q$ has Type II,
$$
N(a,v)=q^{m-2}+\eta( \mu a) q^{m-(r+3)/2}.
$$
\end{lemma}

\section{Linear code from Quadratic form function}

In this manuscript, for $a\in \mathbb{F}_q$, the defining set be defined by
\begin{equation}\label{D-set-a}
D_Q^a =\{ x\in \mathbb{F}_{q^m}: Q(x) = a\}=\{d_1,d_2,\ldots,d_n\}.
\end{equation}
A linear code of length $n$ over $\mathbb{F}_q$ is defined by
\begin{equation}\label{C_D}
\mathcal{C}_{D_Q^a} =\{(Tr(xd_1),Tr(xd_2),\ldots,Tr(xd_n)):x\in\mathbb{F}_{q^m}\}.
\end{equation}
Note that for the case of $a=0$, the weight distribution corresponding linear code has been discussed in \cite{Zhou-li-fan-Helleseth} for $Q(x)$ over $\mathbb{F}_p$ with $r=m$ and in \cite{Zhang-Zhou-peng} for general quadratic form $Q(x)$ over $\mathbb{F}_q$, Thus we only consider the completely weight enumerator of linear code $\mathcal{C}_{D_Q^a}$ for all $a\in \mathbb{F}_q^*.$

\begin{lemma}\label{length-code}  Let $Q(x)$ be a quadratic form of rank $r$ from $\mathbb{F}_{q^m}$ to $\mathbb{F}_q.$ For any $a\in \mathbb{F}_q^*,$ the length of the code $\mathcal{C}_{D_Q^a}$ is
$$|D_Q^a| = q^{m-1} -\epsilon q^{m-r/2-1}  $$
if $r$ is even,
and
$$|D_Q^a| =  q^{m-1}+\eta(\mu a) q^{m-(r+1)/2}$$
otherwise.
\end{lemma}
Proof. One can get the results from Lemma \ref{klapper-qf-1} directly.

\begin{theorem}\label{r-even}
Let $g$ be a generator of $\mathbb{F}_q^*$, if $r$ is even, then the code $\mathcal{C}_{D_Q^a}$ is a $[q^{m-1}-\epsilon q^{m-\frac{r}{2}-1}, m]$ linear code with weight distribution given in Table \ref{r-even-b} and its complete weight enumerator is
\begin{table}
\caption{The weight distribution of the codes $\mathcal{C}_{D_Q^a}$ with $r$  even }
\label{r-even-b}
\begin{center}
\begin{tabular}{|c|c|} \hline
Weight                                                                         & Multiplicity  \\ \hline
0                                                                              & 1 \\ \hline
$(q-1)(q^{m-2}-\epsilon q^{m-\frac{r}{2}-2})$                                  & $q^m-q^r$ \\ \hline
$q^{m-1}-q^{m-2}$                                                              & $\frac{q+1}{2}q^{r-1}+\epsilon \frac{q-1}{2}q^{\frac{r-2}{2}}-1$ \\ \hline
$q^{m-1}-q^{m-2}-2\epsilon q^{m-\frac{r}{2}-1}$                                & $\frac{q-1}{2}(q^{r-1}-\epsilon q^{\frac{r-2}{2}})$\\ \hline
\end{tabular}
\end{center}
\end{table}

\begin{eqnarray*}
&&CWE(\mathcal{C}_{D_Q^a})=\omega_0^{q^{m-1}-\epsilon q^{m-\frac{r}{2}}-1}+(q^m-q^r)\prod\limits_{\rho=0}^{q-1}\omega_\rho^{q^{m-2}-\epsilon q^{m-\frac{r}{2}-2}}\\
&&+(q^{r-1}+\epsilon(q-1)q^{\frac{r}{2}-1}-1)\omega_0^{q^{m-2}-\epsilon q^{m-\frac{r}{2}-1}}\prod\limits_{\rho=1}^{q-1}\omega_\rho^{q^{m-2}}\\
&&+(q^{r-1}-\epsilon q^{\frac{r-2}{2}})\sum\limits_{\beta=1}^{\frac{q-1}{2}}\omega_{0}^{q^{m-2}+\epsilon q^{m-\frac{r}{2}-1}} \omega _{2g^\beta}^{q^{m-2}}\omega_{q-2g^\beta}^{q^{m-2}} \prod\limits_{\rho\neq 0, \pm 2g^\beta} \omega_\rho^{q^{m-2}+(\frac{4g^{2\beta}-\rho^2}{q})\epsilon q^{m-\frac{r}{2}-1}}\\
&&+(q^{r-1}-\epsilon q^{\frac{r}{2}-1})\sum\limits_{\beta=1}^{\frac{q-1}{2}}\omega_{0}^{q^{m-2}-\epsilon q^{m-\frac{r}{2}-1}} \prod\limits_{\rho=1}^{q-1} \omega_\rho^{q^{m-2}+(\frac{4g^{2\beta+1}-\rho^2}{q})\epsilon q^{m-\frac{r}{2}-1}}.
\end{eqnarray*}
\end{theorem}

Proof: By Lemma \ref{length-code}, it is obviously that the code $\mathcal{C}_{D_Q^a}$ has length $n=|D_Q^a|=q^{m-1} -\epsilon q^{m-r/2-1}$ and dimension $m$. For any codeword $\mathbf{c}_b$ in $\mathcal{C}_{D_Q^a}$, according to the definition, its Hamming weight is equal to $$WT(\mathbf{c}_b)=|D_Q^a|-N(a,0).$$ Then, the weight distribution of $\mathcal{C}_{D_Q^a}$ follows from Lemmas \ref{klapper-2}, \ref{klapper-3} and \ref{length-code}.

To obtain the complete weight enumerator of $C_{D_Q^a}$, we need to determine the value
distribution of $\mathcal{N}_b(a,v)$ for each $v\in \mathbb{F}_q$ when $b$ runs through all the elements in $\mathbb{F}_{q^m}$.
Let $\{\alpha_1, \alpha_2,\dots, \alpha_m\} $ and $\{\beta_1, \beta_2,\dots, \beta_m\}$ be the dual basis of $\mathbb{F}_{q^m}$ over $\mathbb{F}_q$. Using the dual bases,
we write $x = x_1\beta_1 +x_2\beta_2 +\ldots+x_m\beta_m$ and $b= b_1\alpha_1 +b_2\alpha_2 +\ldots+b_m\alpha_m$ for $x,~b \in \mathbb{F}_{q^m}$ and we write corresponding vectors as $\bar{x} = (x_1, x_2, \ldots, x_m ) \in \mathbb{F}_q^m$
and $\bar{b} = (b_1, b_2, \ldots, b_m ) \in \mathbb{F}_q^m$. So, the linear function $Tr^m_1(bx)=v $ is equivalent to $L_{\bar{b}}(\bar{x})=v$, and $\mathcal{N}_b(a,v)$ is is equal to the number of solutions $\bar{x} \in \mathbb{F}_q^m \setminus \{ \bar{0} \}$
$$
\left\{
\begin{array}{ll}
\hat{Q}(\bar{x})=a,\\
L_{\bar{b}}(\bar{x})=v.
\end{array}
\right.
$$
after nonsigular transformation for the quadratic form $Q(x)$ (for simplicity, we still use symbol $\bar{x}$).

Observe that $b=0$ gives the zero codeword and the contribution to the complete weight enumerator is $\omega_0^n$, so below we assume that $b\in \mathbb{F}_q^*$.
Note that $\bar{b}$ runs through all the elements in $\mathbb{F}_q^m \setminus \{\bar{0}\}$ if and only if $b$ runs though all the elements in $\mathbb{F}_{q^m}^*.$

Since $g$ is a generator of $\mathbb{F}_q^*$, i.e. each element of $\mathbb{F}_q^*$ can be represented by $g^\beta$ for some $1\le \beta \le q-1$, then when $\beta $ runs over all the element of the set $\{1,2, \ldots, \frac{q-1}{2} \}$, $g^{2\beta}$ and $g^{2\beta+1}$ will run over all the quadratic residue elements and quadratic nonresidue elements of $\mathbb{F}_q^*$, respectively.

The desired conclusions then follow from Lemmas \ref{klapper-2} and \ref{klapper-3}.\hfill
$\Box$

\begin{theorem}\label{r-odd}
Let $g$ be a generator of $\mathbb{F}_q^*$, if $r$ is odd, then the code $\mathcal{C}_{D_Q^a}$ is a $[q^{m-1}+\eta(\mu a) q^{m-\frac{r+1}{2}}, m]$ linear code with weight distribution given in Table \ref{r-odd-b} and its complete weight enumerator is
\begin{table}
\caption{The weight distribution of the codes $\mathcal{C}_{D_Q^a}$ with $r$ odd }
\begin{center}
\label{r-odd-b}
\begin{tabular}{|c|c|}\hline
Weight                                                                          & Multiplicity  \\ \hline
0                                                                               & 1             \\ \hline
$(q-1)\Big(q^{m-2}+\eta(\mu a)q^{m-\frac{r+3}{2}}\Big)$                         & $q^m-q^r+\frac{q-1}{2}\Big(q^{r-1}-\eta(\mu a)q^{\frac{r-1}{2}}\Big)$ \\ \hline
$q^{m-1}-q^{m-2}$                                                               & $q^{r-1}-1$ \\ \hline
$q^{m-2}(q-1)+\eta(\mu a)(q+1)q^{m-\frac{r+3}{2}}$ & $\frac{q-1}{2}\Big(q^{r-1}+\eta(\mu a)q^{\frac{r-1}{2}}\Big)$\\ \hline
\end{tabular}
\end{center}
\end{table}

\begin{eqnarray*}
&&CWE(\mathcal{C}_{D_Q^a})=\omega_0^{q^{m-1}+\eta(\mu a) q^{m-\frac{r+1}{2}}}+(q^m-q^r)\prod\limits_{\rho=0}^{q-1}\omega_\rho^{q^{m-2}+\eta(\mu a)q^{m-\frac{r+3}{2}}}\\
&&+(q^{r-1}-1)\omega_0^{q^{m-2}+\eta(\mu a)q^{m-\frac{r+1}{2}}}\prod\limits_{\rho=1}^{q-1}\omega_\rho^{q^{m-2}}\\
&&+(q^{r-1}+\eta(\mu a)q^{\frac{r-1}{2}}) \sum\limits_{\beta=1}^{\frac{q-1}{2}} \omega_{0}^{q^{m-2}-\eta(\mu a)q^{m-\frac{r+3}{2}}} \omega _{2g^\beta}^{q^{m-2}+(q-1)\eta(\mu a)q^{m-\frac{r+3}{2}}}\\
&&\cdot \omega_{q-2g^\beta}^{q^{m-2}+(q-1)\eta(\mu a)q^{m-\frac{r+3}{2}}} \prod\limits_{\rho\neq 0, \pm 2g^\beta} \omega_\rho^{q^{m-2}-\eta(\mu a)q^{m-\frac{r+3}{2}}}\\
&&+\frac{q-1}{2}(q^{r-1}-\eta(\mu a)q^{\frac{r-1}{2}}) \prod\limits_{\rho=0}^{q-1} \omega_\rho^{q^{m-2}+\eta(\mu a)q^{m-\frac{r+3}{2}}}.
\end{eqnarray*}
\end{theorem}
The proof of  the theorem is very similar to that of Theorem \ref{r-even}, thus we omit it here.

For the quadratic form $Q(x)$ over $\mathbb{F}_q$ with full rank $r=m$, we have following two corollaries corresponding to Theorems \ref{r-even} and \ref{r-odd}, respectively.

\begin{corollary}\label{r-even-c}
Let $Q(x)$ be a quadratic form of full rank from $\mathbb{F}_{q^m}$ to $\mathbb{F}_q$. If $m$ is even, then $\mathcal{C}_{D_Q^a}$ is a two-weight $[q^{m-1}-\epsilon q^{\frac{m}{2}-1}, m]$ code over $\mathbb{F}_q$ with the weight distribution given in Table \ref{r-even-cb} and its complete weight enumerator is
\begin{table}
\caption{The weight distribution of the codes  $\mathcal{C}_{D_Q^a}$ with full rank $m$ even }
\begin{center}
\label{r-even-cb}
\begin{tabular}{|c|c|} \hline
Weight                                                                         & Multiplicity  \\ \hline
0                                                                              & 1 \\ \hline
$q^{m-1}-q^{m-2}$                                                              & $\frac{q+1}{2}q^{m-1}+\epsilon \frac{q-1}{2}q^{\frac{m-2}{2}}-1$ \\ \hline
$q^{m-1}-q^{m-2}-2\epsilon q^{\frac{m}{2}-1}$                                  & $\frac{q-1}{2}(q^{m-1}-\epsilon q^{\frac{m-2}{2}})$\\ \hline
\end{tabular}
\end{center}
\end{table}
\begin{eqnarray*}
&&CWE(\mathcal{C}_{D_Q^a})=\omega_0^{q^{m-1}-\epsilon q^{m-\frac{r}{2}}-1}+(q^{m-1}+\epsilon(q-1)q^{\frac{m}{2}-1}-1)\omega_0^{q^{m-2}-\epsilon q^{\frac{m}{2}-1}}\prod\limits_{\rho=1}^{q-1}\omega_\rho^{q^{m-2}}\\
&&+(q^{m-1}-\epsilon q^{\frac{m-2}{2}})\sum\limits_{\beta=1}^{\frac{q-1}{2}}\omega_{0}^{q^{m-2}+\epsilon q^{\frac{m}{2}-1}} \omega _{2g^\beta}^{q^{m-2}}\omega_{q-2g^\beta}^{q^{m-2}} \prod\limits_{\rho\neq 0, \pm 2g^\beta} \omega_\rho^{q^{m-2}+(\frac{4g^{2\beta}-\rho^2}{q})\epsilon q^{\frac{m}{2}-1}}\\
&&+(q^{m-1}-\epsilon q^{\frac{m}{2}-1})\sum\limits_{\beta=1}^{\frac{q-1}{2}}\omega_{0}^{q^{m-2}-\epsilon q^{\frac{m}{2}-1}} \prod\limits_{\rho=1}^{q-1} \omega_\rho^{q^{m-2}+(\frac{4g^{2\beta+1}-\rho^2}{q})\epsilon q^{\frac{m}{2}-1}}.
\end{eqnarray*}
\end{corollary}

\begin{corollary}\label{r-odd-c}
Let $Q(x)$ be a quadratic form of full rank from $\mathbb{F}_{q^m}$ to $\mathbb{F}_q$. If $m$ is odd, then $\mathcal{C}_{D_Q^a}$ is a three-weight $[q^{m-1}+\eta(\mu a) q^{\frac{m-1}{2}}, m]$ code over $\mathbb{F}_q$ with the weight distribution given in Table \ref{r-odd-cb} and its complete weight enumerator is
\begin{table}
\caption{The weight distribution of the codes $\mathcal{C}_{D_Q^a}$ with full rank $m$ odd }
\begin{center}
\label{r-odd-cb}
\begin{tabular}{|c|c|}\hline
Weight                                                                          & Multiplicity  \\ \hline
0                                                                               & 1             \\ \hline
$q^{m-1}-q^{m-2}$                                                               & $q^{m-1}-1$ \\ \hline
$q^{m-2}(q-1)+\eta(\mu a)(q+1)q^{\frac{m-3}{2}}$                                & $\frac{q-1}{2}\Big(q^{m-1}+\eta(\mu a)q^{\frac{m-1}{2}}\Big)$\\ \hline
$(q-1)\Big(q^{m-2}+\eta(\mu a)q^{\frac{m-3}{2}}\Big)$                             & $\frac{q-1}{2}\Big(q^{m-1}-\eta(\mu a)q^{\frac{m-1}{2}}\Big)$\\ \hline
\end{tabular}
\end{center}
\end{table}

\begin{eqnarray*}
&&CWE(\mathcal{C}_{D_Q^a})=\omega_0^{q^{m-1}+\eta(\mu a) q^{\frac{m-1}{2}}}+(q^{m-1}-1)\omega_0^{q^{m-2}+\eta(\mu a)q^{\frac{m-1}{2}}}\prod\limits_{\rho=1}^{q-1}\omega_\rho^{q^{m-2}}\\
&&+(q^{m-1}+\eta(\mu a)q^{\frac{m-1}{2}}) \sum\limits_{\beta=1}^{\frac{q-1}{2}} \omega_{0}^{q^{m-2}-\eta(\mu a)q^{\frac{m-3}{2}}} \omega _{2g^\beta}^{q^{m-2}+(q-1)\eta(\mu a)q^{\frac{m-3}{2}}}\\
&&\cdot \omega_{q-2g^\beta}^{q^{m-2}+(q-1)\eta(\mu a)q^{\frac{m-3}{2}}} \prod\limits_{\rho\neq 0, \pm 2g^\beta} \omega_\rho^{q^{m-2}-\eta(\mu a)q^{\frac{m-3}{2}}}\\
&&+\frac{q-1}{2}(q^{m-1}-\eta(\mu a)q^{\frac{m-1}{2}}) \prod\limits_{\rho=0}^{q-1} \omega_\rho^{q^{m-2}+\eta(\mu a)q^{\frac{m-3}{2}}}.
\end{eqnarray*}
\end{corollary}

\section{Minimal codewords in $\mathcal{C}_{D_Q^a}$}

In this section, we will show that for most of quadratic form $Q(x)$ and $q$ and $a$, each of codewords of $\mathcal{C}_{D_Q^a}$ given by (\ref{D-set-a}) is minimal,
before doing this, we need to introduce some definitions \cite{Ding-Ding}.

The support of a vector $\mathbf{c }= (c_0,\ldots, c_{n-1}) \in \mathbb{F}_q^n$ is defined as
$$\{0 \le i \le n-1 : c_i \not = 0\}.$$
We say that a vector $\mathbf{x}$ covers a vector $\mathbf{y}$ if the support of $\mathbf{x}$ contains that of
$\mathbf{x}$ as a proper subset.

A minimal codeword of a linear code $\mathcal{C}$ is a nonzero codeword that dose
not cover any other nonzero codeword of $\mathcal{C}$. It is an interesting problem to construct
codes whose nonzero codewords are all minimal since such
linear codes can be employed to construct secret sharing
schemes with interesting access structures \cite{Yuan-Ding-05}. For minimal codewords, we have following results \cite{AB-98,ABCH-95}:
\begin{lemma}\label{lem-min-lc}
In an $[n,k,d]$ code $\mathcal{C}$, let $w_{min}$ and $w_{max}$  be the minimum and maximum nonzero weights, respectively. If
$$\frac{w_{min}}{w_{max}} >\frac{q-1}{q},$$ then all nonzero codewords of $\mathcal{C}$ are minimal.
\end{lemma}

For the code $\mathcal{C}_{D_Q^a}$ given by (\ref{D-set-a}) with $r$ even, if $Q(x)$ is equivalent to Type I, we have
 $$\frac{w_{min}}{w_{max}}=\frac{q^{m-1}-q^{m-2}-2q^{m-\frac{r}{2}-1}}{q^{m-1}-q^{m-2}}>\frac{q-1}{q}$$
 holds if $r=4$ and $q\geq 5,$ or if even $r\ge 6$. While $Q(x)$ is equivalent to Type III, we have
  $$\frac{w_{min}}{w_{max}}=\frac{q^{m-1}-q^{m-2}}{q^{m-1}-q^{m-2}+2q^{m-\frac{r}{2}-1}}>\frac{q-1}{q}$$ holds for
all even $r\geq 4$.

Similarly, for the code $\mathcal{C}_{D_Q^a}$ given by (\ref{D-set-a}) with $r$ odd, we have $\frac{w_{min}}{w_{max}}>\frac{q-1}{q}$ holds if $r\geq 5$.

By Lemma \ref{lem-min-lc}, for certain smaller $r$ and if $m \geq r \ge 5$ for all odd prime power $q$, the linear codes $\mathcal{C}_{D_Q^a}$ of this paper satisfy the condition $\frac{w_{min}}{w_{max}}>\frac{q-1}{q}$, and can be employed to obtain secret sharing schemes with nice access structures. We omit the details here since it is similar to that of \cite{Ding-Ding}.

\section{Conclusion}
In this paper, for any $a\in \mathbb{F}_q^*, $ we constructed some classes of linear code based on the preimage of the general quadratic functions over extension field $\mathbb{F}_q$ and discussed their complete weight enumerators. Results shows that the complete weight enumerators and thus their weight distribution depend on both the parity of rank $r$ and the symbol of $\eta(a)$. More precisely, when $r$ is even, the weight distribution of the depend only on the parity of rank $r$ and posses two weight distribution or three weight for general quadratic form functions or quadratic form functions with $r=m$, respectively.
While $r$ is odd, the weight distribution of the depend on the parity of rank $r$ and posses three weight for all quadratic form functions. Moreover, we obtain that all the nonzero codewords of $\mathcal{C}_{D_Q^a}$ are minimal if $m \geq r \ge 5$, and thus can be employed to obtain secret sharing schemes with interesting access structures. Our results may extended all the corresponding construction from the quadratic form functions.

\end{document}